\documentclass[english,twocolumn]{revtex4}
\usepackage[T1]{fontenc}
\usepackage[latin9]{inputenc}
\usepackage{float}
\usepackage{amsthm}
\usepackage{bm}
\usepackage{amsmath}
\usepackage{graphicx}
\usepackage{amssymb}
\usepackage{esint}

\providecommand{\tabularnewline}{\\}

\theoremstyle{plain}

\usepackage{babel}

\begin{document}

\title{Electronic Coherence Dephasing in Excitonic Molecular Complexes:
Role of Markov and Secular Approximations}

\author{Jan Ol\v{s}ina and Tom\'{a}\v{s} Man\v{c}al}

\affiliation{Institute of Physics of Charles University, Faculty of Mathematics
and Physics, Charles University in Prague, Ke Karlovu 5, CZ-121 16
Prague 2, Czech Republic}
\begin{abstract}
We compare four different types of equations of motion for reduced
density matrix of a system of molecular excitons interacting with
thermodynamic bath. All four equations are of second order in the
linear system-bath interaction Hamiltonian, with different approximations
applied in their derivation. In particular we compare time-nonlocal
equations obtained from so-called Nakajima-Zwanzig identity and the
time-local equations resulting from the partial ordering prescription
of the cummulant expansion. In each of these equations we alternatively
apply secular approximation to decouple population and coherence dynamics
from each other. We focus on the dynamics of intraband electronic
coherences of the excitonic system which can be traced by coherent
two-dimensional spectroscopy. We discuss the applicability of the
four relaxation theories to simulations of population and coherence
dynamics, and identify features of the two-dimensional coherent spectrum
that allow us to distinguish time-nonlocal effects.
\end{abstract}



\email{tomas.mancal@mff.cuni.cz}

\maketitle

\section{Introduction}

Modeling molecular properties related to their non-equilibrium dynamics
requires various theoretical approaches depending on the particular
microscopic processes related to the observed molecular features.
Since the dawn of quantum mechanics, properties of molecules and solids
have been studied theoretically in ever greater detail. This has led
in recent years to a state in which dynamics of complex systems with
multitude of degrees of freedom (DOF) is accessible to quantitative
theoretical study \cite{ComplexMolSystem}. Many properties of molecular
systems are directly related to the equilibrium or time dependent
conformations of nuclear DOF for which electronic states play the
role of a background contributing to the nuclear potential energy
surfaces. Problems like these are the realm of molecular dynamics
(MD) in its classical, quantum or mixed versions and quantum chemistry
(QC), where impressive qualitative and quantitative results have been
achieved in recent years. For certain types of dynamical problems,
however, less expensive model approaches are the preferred choice
due to the scale of studied system or due to the physical nature of
studied processes. A good example of such a problem is ultrafast photo-induced
excited state dynamics of small molecular systems and their aggregates
\cite{PhotosynthExcitonsBook}. Here, most of the relevant experimental
information is only available through ultrafast non-linear spectroscopy,
and thus the theory has to span the whole distance between the microscopic
dynamics of the molecular system, and the macroscopic description
of experimental signals \cite{MukamelBook}. Typical field in which
such an approach has yielded deep understanding of the relevant physico-chemical
processes is the study of primary processes in photosynthesis. The
related quantum mechanical problem is usually formulated in terms
of a model describing the relevant DOF of the system (electronic states
of photosynthetic molecules), and a thermodynamics bath (the protein
environment). Parameters for such models can be supplied by experiment
\cite{Renger01}, QC studies \cite{Scholes01,Madjet01}, MD modeling
\cite{Schulten01}, or are a result of suitable simplified models
\cite{MukamelBook}. 

Recent advances in non-linear spectroscopy have opened a wide new
experimental window into the details of ultrafast photo-induced dynamics
of molecular systems. Experimental realization of two-dimensional
(2D) coherent spectroscopy in the visible and near IR regions \cite{Jonas01,Miller01,BrixnerStiopkin,BrixnerMancal}
has enabled to overcome some of the frequency- vs. time-resolution
competition problem otherwise faced by ultrafast spectroscopy, and
yielded thus unprecedented experimental details of the time evolution
of molecular excitations. Most importantly, it was predicted that
the presence of certain oscillatory features in 2D spectra is a manifestation
of coherences between molecular excited states \cite{PisliakovMancal,Bruggemann01}.
It was also concluded that these oscillations should be present in
the 2D spectrum of photosynthetic Fenna-Matthews-Olson (FMO) chromophore-protein
complex \cite{PisliakovMancal}. Experimental results not only confirmed
this prediction \cite{EngelNature}, but yielded also surprising results
such as unexpectedly long life time of these coherences, as compared
to the predictions of standard dephasing rate theory. Furthermore,
while possible coherence transfer was ignored by the relaxation theory
used in Ref. \cite{PisliakovMancal}, the experiment provided some
evidence for its role in excitation energy transfer. It was speculated
that photosynthetic systems might use the coherent mode of energy
transfer to more efficiently channel excitation energy by scanning
their energetic landscape in a process similar to quantum computing
\cite{EngelNature}. More experiments have recently reported coherent
dynamics in photosynthetic systems \cite{Lee_Science} and conjugated
polymers \cite{Scholes02}, and the field of energy transfer in photosynthesis
has seen an increased interest from theoretical researchers from previously
unrelated fields \cite{Rebentrost,Mohseni,Plenio,Olaya-Castro}.

Theoretical basis for the description of the decoherence phenomena
in excitation energy transfer has been developed long ago in the framework
of the reduced density matrix (RDM) \cite{Nakajima,Zwanzig}. Equations
of motion resulting from this scheme are characterized by the presence
of time retarded terms responsible for energy relaxation and decoherence
processes. Equations of this type will be denoted as \emph{time non-local}
in this paper. Later, an alternative approach to the derivation of
the RDM equations of motion has emerged which yields \emph{time local}
equation of motion \cite{Hashitsume,Shibata}. Both theories express
the relaxation term in form of an infinite series in terms of the
system-bath interaction Hamiltonian, but differ in time ordering prescriptions
for the cumulant expansion of the evolution operator. The time local
equations correspond to so-called partial time ordering prescription
of the cummulant expansion, while the time non-local equations result
from so-called chronological time ordering \cite{Mukamel1,Mukamel2}.
Although the two schemes yield formally different equations of motion
for the RDM, they are in fact equivalent as long as the complete summation
of the corresponding infinite series is performed. When the infinite
series are truncated at a finite order, the two theories yield equations
that predict different RDM dynamics. This is a result of different
statistical assumptions about the bath that are implicitly made in
the two cases \cite{Mukamel1,Mukamel2}. In all orders of expansion,
so-called \emph{Markov approximation} can be used to transform the
time non-local equation of motion into a certain time-local form.
This has to be regarded as an additional approximation which simplifies
the numerical treatment of the time non-local equations. Interestingly,
in the second order the time-local equations and the time non-local
equations with Markov approximation have exactly the same form. 

Until recently, most experiments were not sensitive to coherence between
electronic levels. This allowed a host of further approximations to
simplify equations of motion. Most notably, the \emph{secular approximation},
which amounts to decoupling RDM elements oscillating on different
frequencies from each other, has limited the energy transfer phenomena
to separate dynamics of population transfer and coherence dephasing
\cite{MayBook}. Even on very short times scale, experiments aimed
at studying population dynamics (pump probe) are not sensitive enough
to coherences between electronic levels to require non-secular theory,
although it was suggested that measured relaxation time can be distorted
by non-secular effects \cite{Capek}. Consequently, most of the theory
developed for evaluation of experiments has been aimed at improving
calculation of the relaxation rates \cite{Zhang,Mino,Silbey04}. With
experiments now uncovering new details about the role of electronic
coherence, theoretical methods beyond rate equations for probabilities
which are both accurate and numerically tractable are required. Although
schemes for constructing equations of motion for the RDM beyond second
order, based on co-propagation of the RDM with auxiliary operators,
seem feasible and promising \cite{Ishizaki1,Ishizaki2}, second order
theories might still be the only option for treatment of extended
molecular systems. It was suggested previously that second order perturbation
theory with respect to system-bath coupling provides a suitable framework
for development of such methods \cite{Mancal08a}. This notion is
also supported by the fact that in the special case of so-called spin-boson
model, second order time-local equation of motion already represents
an exact equation of motion for the RDM \cite{Haenggi08}. 

In this paper we study the following four different second order theories:
(a) full time non-local (full TNL) equation of motion resulting from
the Nakajima-Zwanzig identity or equivalently from the chronological
ordering prescription in the cummulant expansion, (b) the full time
local (full TL) equation of motion resulting from the partial ordering
prescription in the cummulant expansion, or equivalently from Markov
approximation applied to TNL equation, (c) time non-local equation
with secular approximation (secular TNL), and (d) time local equation
with secular approximation (secular TL). We discuss the applicability
of these equations to the description of the energy relaxation and
decoherence dynamics in small systems of molecular excitons with the
emphasis of on recent 2D spectroscopic experiments and the dynamics
of coherence between electronic excited states. Note that, in this
paper, \emph{full }refers to the equations where no secular approximation
has been applied. These equations are still of second order of perturbation
theory.

The paper is organized as follows. The next section introduces Hamiltonian
description of an aggregate of small molecules embedded in a protein
or solid state environment. In Section \ref{sec:Second-order-relaxation}
we describe the details of second order theory of system--bath interaction
and we derive four different equations of motion for the RDM describing
electronic states of a molecular aggregate. Two-dimensional coherent
spectroscopy, and non-linear spectroscopy in general are introduced
in Section \ref{sec:Non-linear-Spectroscopic-Signals}. In Section
\ref{sec:Numerical-Results} we present and discuss numerical results
comparing different theories of relaxation on calculations of coherence
life time and 2D spectra.

\section{Model Hamiltonian\label{sec:Model-Hamiltonian}}

The investigated molecular system is an aggregate composed of $N$
monomers embedded in protein environment. Let us first consider a
monomeric molecule (a chromophore) embedded in the environment, but
insulated from interaction with its neighboring monomers. The monomer
Hamiltonian has a form\[
H^{m}=\left(\varepsilon_{g}^{(m)}+T(P_{m})+V_{g}^{(m)}(Q_{m})\right)|g_{m}\rangle\langle g_{m}|\]

\begin{equation}
+\left(\varepsilon_{e}^{(m)}+T(P_{m})+V_{e}^{(m)}(Q_{m})\right)|e_{m}\rangle\langle e_{m}|\;,\label{eq:MonomHam}\end{equation}
where $|g_{m}\rangle$, $|e_{m}\rangle$ denote electronic ground
and excited states, and $\varepsilon_{g}^{(m)}$, $\varepsilon_{e}^{(m)}$
represent electronic energies of these states. The kinetic term $T(P_{m})$
and the potential terms $V_{g}^{(m)}(Q_{m})$, $V_{e}^{(m)}(Q_{m})$
represent the intra-molecular DOF and the protein environment (bath
or reservoir) interacting with these states. By $Q_{m}$ $(P_{m})$
we denote the (possibly macroscopic) set of coordinates (impulses)
describing both the intramolecular nuclear DOF of the $m$-th monomer
as well as the DOF of its surroundings. The total Hamiltonian of the
monomer can be split  into the system, reservoir and the system-reservoir
coupling terms \[
H_{S}^{m}\equiv\varepsilon_{g}^{(m)}|g_{m}\rangle\langle g_{m}|\]
\begin{equation}
+(\varepsilon_{e}^{(m)}+\langle V_{e}^{(m)}(Q_{m})-V_{g}^{(m)}(Q_{m})\rangle_{eq})|e_{m}\rangle\langle e_{m}|\;,\label{eq:MonomHamS}\end{equation}
\[
H_{\mathrm{R}}^{m}\equiv[T(P_{m})+V_{g}^{(m)}(Q_{m})]\]
\[
\times\left(|g_{m}\rangle\langle g_{m}|+|e_{m}\rangle\langle e_{m}|\right)\]
\begin{equation}
=[T(P_{m})+V_{g}^{(m)}(Q_{m})]\otimes\hat{1},\label{eq:MonomHamR}\end{equation}
\[
H_{\mathrm{S-R}}^{m}\equiv\Big(V_{e}^{(m)}(Q_{m})-V_{g}^{(m)}(Q_{m})\]
\[
-\langle V_{e}^{(m)}(Q_{m})-V_{g}^{(m)}(Q_{m})\rangle_{eq}\Big)|e_{m}\rangle\langle e_{m}|\]
\begin{equation}
\equiv\Delta\Phi_{m}(Q_{m})|e_{m}\rangle\langle e_{m}|\;.\label{eq:MonomHamS-R}\end{equation}

Here, $\langle A(Q)\rangle_{eq}$ represents averaging of an arbitrary
$Q-$dependent operator over equilibrium state of the bath. By this
choice of the splitting we have assured that $\Delta\Phi^{(m)}(Q_{m})=0$
for the system in equilibrium. To simplify the notation, we redefine
electronic energy of the excited state to include the equilibrium
average of the potential energy difference between the electronic
excited and ground states, $\tilde{\varepsilon}_{e}^{(m)}=\varepsilon_{e}^{(m)}+\langle V_{e}^{(m)}(Q_{m})-V_{g}^{(m)}(Q_{m})\rangle_{eq}$
and we drop the tilde over $\varepsilon_{e}^{(m)}$ further in this
paper. 

\begin{figure}
\includegraphics[width=1\columnwidth]{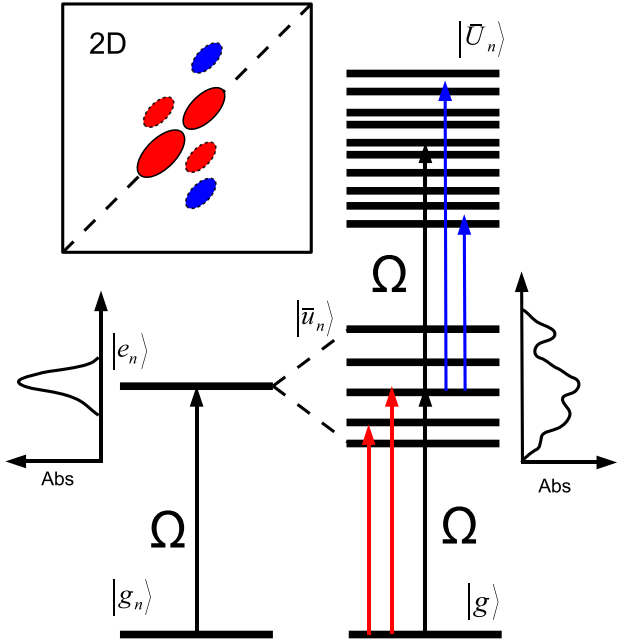}

\caption{\label{fig:Exciton_states}Illustration of the level structure of
an excitonic system. The excited states $|e_{n}\rangle$ of $N$ monomers
with transition frequency $\Omega$ (left part of the figure) split
due to the resonance interaction into $N$ one exciton states $|\bar{u}_{n}\rangle$
(right). Absorption spectra of ensembles of non-interacting (left)
and interaction (right) monomers. The system also exhibits higher
excitons states (two-exciton states $|\bar{U}_{n}\rangle$ are depicted
here), with $\Omega$ being the mean transition frequency from the
one- to two-exciton bands. A pictorial 2D spectrum with peaks resulting
from transitions between the ground- and one-exciton states (red arrows)
and one- and two-exciton states (blue arrows) is presented in the
upper left corner of the figure. The transitions between the ground-
and one-excitons states lead to positive contributions to the 2D spectrum
(absorption and ground state bleach), while the transitions between
the one- and two-exciton states result in a negative contribution
(excited state absorption).}

\end{figure}

An aggregate built out of these monomers can be represented on a Hilbert
space composed of collective aggregate states. We define the aggregate
ground state\begin{equation}
|g\rangle=\prod_{m=1}^{N}\otimes|g_{m}\rangle,\label{eq:col_ground}\end{equation}
states with a single excitation \begin{equation}
|u_{n}\rangle=\prod_{m=1}^{n-1}\otimes|g_{m}\rangle\otimes|e_{n}\rangle\prod_{m^{\prime}=n+1}^{N}\otimes|g_{m^{\prime}}\rangle,\label{eq:coll_one_ex}\end{equation}
and multi-excited states in an analogical manner. We drop the sign
$\otimes$ in further consideration for the sake of brevity. The Hamiltonian
of the aggregate is constructed using the energies of collective states\[
H_{S}^{non-int}=\varepsilon_{g}|g\rangle\langle g|\]
\begin{equation}
+\sum_{n}\left(\Delta\varepsilon_{n}+\Omega\right)|u_{n}\rangle\langle u_{n}|+\mathrm{h.\; e.\; t.},\label{eq:AggHam1Excitonic_a}\end{equation}
where $\varepsilon_{g}=\sum_{n}\varepsilon_{g}^{(n)}$, $\Omega=N^{-1}\sum_{n}\varepsilon_{e}^{(n)}$,
and $\Delta\varepsilon_{n}=\varepsilon_{e}^{(n)}-\Omega+\sum_{m\neq n}\varepsilon_{g}^{(m)}$.
The abreviation h. e. t. denotes higher excitonic terms. Due to the
fact that the monomers are positioned in a tight aggregate, we have
to account for the interaction energy between their excited states.
The interaction energy between states $|u_{m}\rangle$ and $|u_{n}\rangle$
will be denoted $J_{mn}$, and the corresponding contribution to the
total Hamiltonian reads

\begin{equation}
H_{S}^{int}=\sum_{n\neq m}\left(J_{nm}|u_{n}\rangle\langle u_{m}|+\mathrm{c.c}.\right)+\mathrm{h.\; e.\; t.}.\label{eq:AggHam1Excitonic}\end{equation}
 Due to the off-diagonal terms $J_{mn}$ the collective states defined
in Eq. (\ref{eq:coll_one_ex}) are not eigenstates of the total Hamiltonian
$H_{S}=H_{S}^{non-int}+H_{S}^{int}$. Although the basis of the states
$|g\rangle$, $|u_{n}\rangle$ and multiple-excitation states of the
aggregate provides efficient means for defining the Hamiltonian, it
is more practical to switch into the basis of eigenstates of the Hamiltonian
$H_{S}$. One of the reasons is that while matter interacts with light,
the differences between eigenenergies of $H_{S}$ define the resonant
transition frequencies. The system--bath coupling part of the aggregate
Hamiltonian reads\begin{equation}
H_{S-B}=\sum_{n}\Delta\Phi_{n}|u_{n}\rangle\langle u_{n}|+\mathrm{h.\; e.\; t.},\label{eq:H_S-B_agg}\end{equation}
and thus no terms in the total Hamiltonian couple the ground state
with the first excited state or higher excited state bands. In fact,
the total Hamiltonian splits into blocks separated approximately by
the energy $\hbar\Omega$ (see Fig. \ref{fig:Exciton_states}). This
reflects the neglecting of all adiabatic couplings, which are supposed
to be so weak that they do not lead to transitions on the time scale
of interest (femto and picoseconds). This property is well justified
e.g. for chlorophyll systems. 

For the subsequent use in this paper, we denote the eigenstates of
the total electronic Hamiltonian $H_{S}$ by $|\bar{u}_{a}\rangle$,
$a=1,\dots,N$, for single exciton states formed as linear combinations
of single excitation states $|u_{n}\rangle$ and $|\bar{U}_{a}\rangle$,
$a=N+1,\dots,N+N(N-1)/2$, for two-exciton states formed from the
linear combination of pairs of single excitation states.

\section{Second order relaxation theories\label{sec:Second-order-relaxation}}

In this section we consider interaction of the electronic system described
by the Hamiltonian $H_{S}$ with a macroscopic bath composed of the
DOF of the molecular surroundings.

\subsection{Projection operator technique and Nakajima-Zwanzig identity\label{sub:Projection-operator-technique}}

We start with the density operator $W$, which describes the state
of the system composed of the aggregate and its surroundings. This
operator fulfills Liouville-von Neumann equation \begin{equation}
\frac{\partial}{\partial t}W(t)=-\frac{i}{\hbar}[H,W(t)]_{-}=-i{\cal L}W(t).\label{eq:LvN}\end{equation}
Here, $[A,B]_{-}\equiv AB-BA$, and we introduced Liouville superoperator
${\cal L}$ by its action on an arbitrary operator $A$ as ${\cal L}A=\frac{1}{\hbar}[H,A]_{-}$.
Liouville superoperators ${\cal L}_{S}$, ${\cal L}_{B}$ and ${\cal L}_{S-B}$
that correspond to Hamiltonian operators of the system, the bath and
their interaction will be used later in this paper. A formal solution
of Eq. (\ref{eq:LvN}) can be written using evolution superoperators
such that\begin{equation}
W(t)={\cal U}(t-t_{0})W(t_{0}).\label{eq:supop}\end{equation}
We are often not interested in the complete information carried by
$W(t)$, but only in the information related to the electronic states
of the aggregate. The reduced density matrix (RDM) defined as\begin{equation}
\rho(t)=tr_{B}\{W(t)\},\label{eq:RDM}\end{equation}
where $tr_{B}$ represents trace over all bath DOF, is the quantity
carrying just the relevant information \cite{MayBook}. To derive
equation of motion for $\rho(t)$ we can use projection operator approach
as following. We define a projection operator ${\cal P}$ by its action
on an arbitrary operator $A$ as\begin{equation}
{\cal P}A=tr_{B}\{A\}w,\label{eq:P_general}\end{equation}
where $w$ is a pure bath operator satisfying the condition $tr_{B}\{w\}=1$.
For convenience we also define a complementary projection operator
${\cal Q}\equiv1-{\cal P}$. For the projected density matrix ${\cal P}W(t)$
one can write a formally exact equation of motion as\[
\frac{\partial}{\partial t}{\cal P}W(t)=-i{\cal P}{\cal L}_{I}(t){\cal P}W(t)\]
 \[
-{\cal P}{\cal L}_{I}(t)\exp\left\{ -i\int_{t_{0}}^{t}d\tau{\cal Q}{\cal L}_{I}(\tau){\cal Q}\right\} {\cal Q}W(t_{0})\]
\[
-\int\limits _{0}^{t-t_{0}}d\tau^{\prime}{\cal P}{\cal L}_{I}(t)\exp\left\{ -i\int_{t_{0}}^{\tau^{\prime}}d\tau^{\prime\prime}{\cal Q}{\cal L}_{I}(\tau^{\prime\prime}){\cal Q}\right\} \]
\begin{equation}
\times{\cal Q}{\cal L}_{I}(t-\tau^{\prime}){\cal P}W(t-\tau^{\prime}),\label{eq:Naka-Zw}\end{equation}
where the subscript $_{I}$ denotes interaction picture with respect
to the bath Hamiltonian $H_{B}$. Eq. (\ref{eq:Naka-Zw}) is known
as Nakajima-Zwanzig identity \cite{MayBook,FainBook}. Although one
cannot practically solve Eq. (\ref{eq:Naka-Zw}), because it is as
difficult as the original complete problem, one can use it as a starting
point for approximations that will follow.

\subsection{Convolutionless equation of motion\label{sub:Convolutionless-equation}}

Before we introduce approximations into Eq. (\ref{eq:Naka-Zw}), we
can make one more formal step, which enables us to overcome its time
non-local character. Let us assume an evolution superoperator ${\cal U}^{\prime}(t)$,
which satisfies\begin{equation}
{\cal P}W(t)={\cal U}^{\prime}(t-t_{0}){\cal P}W(t_{0}).\label{eq:global_evol}\end{equation}
This superoperator is of course unknown, but it allows us to turn
a time non-local differential equation (\ref{eq:Naka-Zw}) formally
into a time-local one. We can write\[
\frac{\partial}{\partial t}{\cal P}W(t)=-i{\cal P}{\cal L}_{I}(t){\cal P}W(t)\]
\[
-{\cal P}{\cal L}_{I}(t)\exp\left\{ -i\int_{t_{0}}^{t}d\tau{\cal Q}{\cal L}_{I}(\tau){\cal Q}\right\} {\cal Q}W(t_{0})\]
\[
-\Big[\int\limits _{0}^{t-t_{0}}d\tau^{\prime}{\cal P}{\cal L}_{I}(t)\exp\left\{ -i\int_{t_{0}}^{\tau^{\prime}}d\tau^{\prime\prime}{\cal Q}{\cal L}_{I}(\tau^{\prime\prime}){\cal Q}\right\} \]
\begin{equation}
\times{\cal Q}{\cal L}_{I}(t-\tau^{\prime}){\cal U}^{\prime}(-\tau^{\prime})\Big]{\cal P}W(t).\label{eq:Covless}\end{equation}
We do not develop the convolutionless (or time local) theory any further
in this paper, because we will be interested only in terms up to the
second order in ${\cal L}_{I}$. In such a case ${\cal U}^{\prime}(-\tau)$
can be approximated as ${\cal U}^{\prime}(-\tau)\approx{\cal U}_{S}(-\tau){\cal U}_{B}(-\tau)$,
where ${\cal U}_{S}(-\tau)$ and ${\cal U}_{B}(-\tau)$ are evolution
superoperators with respect to ${\cal L}_{S}$ and ${\cal L}_{B}$.
Interested reader can refer e.g. to Refs. \cite{FainBook,Mukamel1,Mukamel2}
for further details.

\subsection{System-Bath Coupling\label{sub:Bilinear-System-Bath-Coupling}}

We will now assume the interaction Hamiltonian in a form of Eq. (\ref{eq:H_S-B_agg})
where index $n$ now runs through all relevant single-exciton and
multi-exciton states \begin{equation}
H_{I}=\sum_{n}\Delta\Phi_{n}K_{n}.\label{eq:bicoupling}\end{equation}
Correspondingly, $K_{n}=|u_{n}\rangle\langle u_{n}|$ for single excitonic
states. Expanding Eqs. (\ref{eq:Naka-Zw}) and (\ref{eq:Covless})
into the second order in $H_{I}$ we find that the third term on the
right hand side (r. h. s.) can be conveniently expressed via so-called
bath (or energy gap) correlation functions defined as\begin{equation}
C_{mn}(\tau)=tr_{Q}\{U_{B}(-\tau)\Delta\Phi_{m}U_{B}(\tau)\Delta\Phi_{n}w_{eq}\}.\label{eq:bathCf}\end{equation}
Here, we have chosen a specific form of the bath density matrix $w\equiv w_{eq}$,
where $w_{eq}$ is the canonical density matrix of the bath DOF. Defining
also an operator \begin{equation}
\Lambda_{m}(\tau)=\sum_{n}C_{mn}(\tau)U_{S}(\tau)K_{n}U_{S}^{\dagger}(\tau)\label{eq:Lambda_m}\end{equation}
and a superoperator ${\cal M}^{(2)}(\tau)$ such that\[
{\cal M}^{(2)}(\tau)A=\sum_{m}[K_{m},\Lambda_{m}(\tau)U_{S}(\tau)AU_{S}(-\tau)\]
\begin{equation}
-U_{S}(\tau)AU_{S}(-\tau)\Lambda_{m}^{\dagger}(\tau)]_{-},\label{eq:M2_in_Lambda}\end{equation}
the Eqs. (\ref{eq:Naka-Zw}) and (\ref{eq:Covless}) can be rewritten
as\[
\frac{\partial}{\partial t}\rho(t)=-i{\cal L}_{S}\rho(t)\]

\begin{equation}
-\sum_{m}\int\limits _{0}^{t-t_{0}}d\tau[K_{m},\Lambda_{m}(\tau)\rho(t)-\rho(t)\Lambda_{m}^{\dagger}(\tau)]_{-}\label{eq:rdm_eq_convless}\end{equation}
and\[
\frac{\partial}{\partial t}\rho(t)=-i{\cal L}_{S}\rho(t)\]
\[
-\sum_{m}\int\limits _{0}^{t-t_{0}}d\tau[K_{m},\Lambda_{m}(\tau)U_{S}(\tau)\rho(t-\tau)U_{S}(-\tau)\]
\begin{equation}
-U_{S}(\tau)\rho(t-\tau)U_{S}(-\tau)\Lambda_{m}^{\dagger}(\tau)]_{-}.\label{eq:rdm_eq_conv}\end{equation}
Provided we can supply a model for the correlation function $C_{mn}(\tau)$
we are in position to write down the equations of motion for RDM in
terms of known quantities. The last step necessary to implement these
equations is to represent them in the basis of the eigenstates of
the aggregate Hamiltonian. We define \begin{equation}
\rho_{ab}(t)=\langle\bar{u}_{a}|\rho(t)|\bar{u}_{b}\rangle,\ a,b=1,\dots,N,\label{eq:rho_ab}\end{equation}
\[
\rho_{ab}(t)=\langle\bar{U}_{a}|\rho(t)|\bar{U}_{b}\rangle,\ \]
\begin{equation}
a,b=N+1,\dots,N+N(N-1)/2,\label{eq:rho_AB}\end{equation}
and in a similar manner for matrix elements of other operators and
superoperators. This leads to\begin{equation}
\frac{\partial}{\partial t}\rho_{ab}(t)=-i\omega_{ab}\rho_{ab}(t)-\sum_{cd}{\cal R}_{abcd}(t)\rho_{cd}(t),\label{eq:rho_ab_R}\end{equation}
with ${\cal R}_{abcd}(t)$ being the matrix elements of the superoperator
defined by the r. h. s. of Eq. (\ref{eq:rdm_eq_convless}), and\[
\frac{\partial}{\partial t}\rho_{ab}(t)=-i\omega_{ab}\rho_{ab}(t)\]
 \begin{equation}
-\sum_{cd}\int\limits _{0}^{t-t_{0}}{\cal M}_{abcd}(\tau)\rho_{cd}(t-\tau),\label{eq:rho_ab_M}\end{equation}
with ${\cal M}_{abcd}(\tau)$ the matrix elements of the superoperator
defined by the r. h. s. of Eq. (\ref{eq:rdm_eq_conv}). All the quantities
needed to calculate the matrix elements ${\cal R}_{abcd}(t)$ and
${\cal M}_{abcd}(t)$ are known provided the energy gap correlation
function is know.

\subsection{Energy gap correlation function\label{sub:Energy-gap-correlation-1}}

As a suitable model of the energy gap correlation function we choose
so-called multi-mode Brownian oscillator (BO) \cite{MukamelBook}.
In general, Brownian oscillator model can interpolate between underdamped
intra-molecular DOF and (usually) overdamped bath DOF representing
the immediate surroundings of the molecule. In this paper, we assume
the correlation function of the energy gap of each molecule in the
aggregate to be the same, and independent of neighboring molecules,
i.e.\begin{equation}
C_{ab}(t)=C(t)\delta_{ab}.\label{eq:Cab_delta}\end{equation}
The correlation function $C(t)$ is taken in a form of the overdamped
BO model\[
C(t)=-i\hbar\lambda\Lambda e^{-\Lambda|t|}\;\mathrm{sgn\;}t\]
 \begin{equation}
+\lambda\Lambda\hbar\coth\left(\frac{\beta\hbar\Lambda}{2}\right)e^{-\Lambda|t|}+\frac{4\Lambda\lambda}{\beta}\sum_{n=1}^{\infty}\frac{\nu_{n}e^{-\nu_{n}|t|}}{{\nu_{n}}^{2}-\Lambda^{2}}\;,\label{eq:BO_c}\end{equation}
with\begin{equation}
\nu_{n}\equiv\frac{2\pi n}{\hbar\beta},\;\beta\equiv\frac{1}{k_{\mathrm{B}}{\cal T}},\;\Lambda\equiv\frac{1}{\tau_{\mathrm{c}}}\;.\label{eq:BO_par}\end{equation}
Here, $\lambda$ is the reorganization energy, $\nu_{n}$ are so-called
Matsubara frequencies, $k_{\mathrm{B}}$ is the Boltzmann constant,
${\cal T}$ is the thermodynamic temperature and $\tau_{\mathrm{c}}$
is the so-called bath correlation time. The BO form of the correlation
function satisfies all general constraints put of a correlation function
by thermodynamics \cite{MukamelBook}. Apart of the temperature which
we assume to be ${\cal T}=300$ K in all calculations in this paper,
the BO model is determined by two parameters only; by the reorganization
energy $\lambda$ which is experimentally related to the Stokes shift
$S=2\lambda$ and by the bath correlation time $\tau_{c}$. BO is
a widely used, well physically motivated, but not the only possible
model for the bath correlation function. Implications of other forms
of the correlation function for the RDM dynamics will be studied elsewhere.

\subsection{Secular and constant relaxation rate approximations in the energy
eigenstate basis\label{sub:Secular-approximation}}

Eqs. (\ref{eq:rho_ab_R}) and (\ref{eq:rho_ab_M}) are systems of
coupled (integro-) differential equations for the elements of the
RDM. From the first terms on the r. h. s. we deduce that the element
$\rho_{ab}(t)$ oscillates with a frequency close to $\omega_{ab}$.
It is often justified to assume that two terms oscillating on different
frequencies are independent of each other. For their envelopes $\bar{\rho}_{ab}(t)=e^{i\omega_{ab}t}\rho_{ab}(t)$
we have \begin{equation}
\frac{\partial}{\partial t}\bar{\rho}_{ab}(t)=-\sum_{cd}{\cal R}_{abcd}(t)e^{i(\omega_{ab}-\omega_{cd})t}\bar{\rho}_{cd}(t),\label{eq:sec1}\end{equation}
and integration over time has therefore a relatively smaller contribution
when $\omega_{ab}-\omega_{cd}\neq0$. Neglecting these contributions,
usually termed \emph{secular approximation} \cite{MayBook}, leads
to setting \begin{equation}
{\cal R}_{abcd}(t)=0,\label{eq:sec2}\end{equation}
for all term except when $a=c$ and $b=d$ , or $a=b$ and $c=d$.
The interpretation of the remaining non-zero terms is simple. The
terms ${\cal R}_{aabb}(t)$ for $a\neq b$ represent rates of transition
from level denoted by index $b$ to a level denoted by $a$. The term
${\cal R}_{aaaa}(t)$ corresponds to the total transition rate from
the level $a$ to all other levels. The terms ${\cal R}_{abab}(t)$
($a\neq b)$ are rates of the damping of a coherence element $\rho_{ab}(t)$.
In secular approximation, the dynamics of populations of electronic
levels is thus decoupled from the dephasing of coherences. The above
arguments for the secular approximation apply also to the integro-differential
equation (\ref{eq:rho_ab_M}), and we can thus define four different
second order equations of motion for the RDM, with different levels
of approximation. From the perspective of our derivation, the most
general second order equation is Eq. (\ref{eq:rho_ab_M}), which we
have denoted full TNL. The convolutionless Eq. (\ref{eq:rho_ab_R})
denoted full TL can be regarded its approximation, but it can also
be alternatively viewed as derived by different cummulant approximation,
see Refs. \cite{Mukamel1,Mukamel2}. The set of four methods investigated
here is completed by applying secular approximation to the full TNL
and full TL equations. 

All four sets of equations of motion we consider here are extensions
to the two well-known constant relaxation rate theories. To arrive
at the well-known Redfield equations \cite{MayBook}, one can assume
certain coarse graining of the RDM dynamics so that all significant
changes to the $\rho(t)$ occur on a time scale much longer than the
correlation time $\tau_{c}$. Then time $t_{0}$ in Eq. (\ref{eq:rdm_eq_convless})
can be put to $-\infty$ and the integration limits are then from
zero to infinity. The relaxation tensor ${\cal R}$ thus becomes time
independent. If we, on the other hand, consider the decay of $C(t)$
to be much faster than even the transition frequencies between electronic
levels, we can assume $C(t)\approx C_{0}\delta(t)$ and Eq. (\ref{eq:rdm_eq_convless})
has the well-known Lindblad form \cite{Lindblad76a,MayBook}. Only
for the Lindblad form and for the Redfield equations in secular approximation,
it can be shown that the diagonal elements of $\rho(t)$ are always
positive. For all other equations we have derived here, this assertion
cannot be proven in general. This is a consequence of the fact that
they are derived in a low order of perturbation theory.

\section{Non-linear Spectroscopic Signals\label{sec:Non-linear-Spectroscopic-Signals}}

Non-linear spectroscopic signals are very well described by time-dependent
perturbation theory \cite{MukamelBook}. The equations of motion,
Eqs. (\ref{eq:rdm_eq_convless}) to (\ref{eq:rdm_eq_conv}), can be
extended by semiclassical light-matter interaction term. This yields\begin{equation}
\frac{\partial}{\partial t}\rho(t)=-i{\cal L}_{S}\rho(t)-{\cal D}[\rho(t)](t)+i{\cal V}\rho(t)E(t),\label{eq:EM_with_E}\end{equation}
where $E(t)=\bm{n}\cdot\bm{E}(t)$, is the projection of the external
electric field vector $\bm{E}(t)$ on the normal vector $\bm{n}$
in direction of the molecular transition dipole moment. The symbol
$D[\rho(t)](t)$ represents the relaxation term chosen from the full
TNL, full TL, secular TNL and secular TL equations of motion. The
superoperator ${\cal V}$ is a commutator with the dipole moment operator
$\bm{\mu}=\bm{n}\mu$, so that for an arbitrary operator $A$ we have\begin{equation}
{\cal V}A=\frac{1}{\hbar}[\mu,A]_{-}.\label{eq:V_sup}\end{equation}

\subsection{Third-order non-linear response theory\label{sub:Third-order-non-linear-response}}

Non-linear optical signals are related to the RDM via polarization
\begin{equation}
P(t)=tr\{\mu\rho(t)\}.\label{eq:polarization}\end{equation}
In particular, for the third order non-linear signal $E_{s}^{(3)}(t)$
one can write\begin{equation}
E_{s}^{(3)}(t)\approx i\omega P^{(3)}(t)=i\omega\ tr\{\mu\rho^{(3)}(t)\},\label{eq:sig_from_rho}\end{equation}
where the upper index $^{(3)}$ denotes that the quantity is of the
third order of the perturbation theory with respect to the external
electric field $E(t)$. By defining the evolution superoperator ${\cal U}(t)$
which fulfills Eq. (\ref{eq:EM_with_E}) with $E(t)=0$ we can write
the third order perturbation term as\[
\rho^{(3)}(t)=-i\int\limits _{0}^{\infty}\int\limits _{0}^{\infty}\int\limits _{0}^{\infty}d\tau_{3}d\tau_{2}d\tau_{1}{\cal U}(\tau_{3}){\cal V}{\cal U}(\tau_{2}){\cal V}{\cal U}(\tau_{1}){\cal V}\rho_{0}\]
\begin{equation}
\times E(t-\tau_{3})E(t-\tau_{3}-\tau_{2})E(t-\tau_{3}-\tau_{2}-\tau_{1}).\label{eq:rho3}\end{equation}
In experiment, the laser field is often prepared in a form of three
incident pulses\[
E(t)=A_{1}(t-t_{1})e^{-i\Omega_{1}(t-t_{1})+i\bm{k}_{1}\bm{r}}\]
\[
+A_{2}(t-t_{2})e^{-i\Omega_{2}(t-t_{2})+i\bm{k}_{2}\bm{r}}\]
\begin{equation}
+A_{3}(t-t_{3})e^{-i\Omega_{3}(t-t_{3})+i\bm{k}_{3}\bm{r}}+c.c.,\label{eq:Et_in_A}\end{equation}
with different k-vectors $\bm{k}_{1}$, $\bm{k}_{2}$ and $\bm{k}_{3}$.
In the rest of the paper we assume $\Omega_{1}=\Omega_{2}=\Omega_{3}\equiv\Omega$,
$A_{1}(t)=A_{2}(t)=A_{3}(t)\equiv A(t)$. The expression obtained
by inserting Eq. (\ref{eq:rho3}) into Eq. (\ref{eq:sig_from_rho})
can be significantly simplified in cases where the system consists
of a ground-state and a band of excited states, with the transition
frequency close to resonance with the laser pulse frequency $\Omega$,
and by assuming the laser pulses are ultra short, i.e. $A(t)\approx E_{0}\delta(t)$.
For an experiment which detects non-linear signal emitted in the direction
$-\bm{k}_{1}+\bm{k}_{2}+\bm{k}_{3}$, the third order signal has a
frequency $\approx\Omega$ and it is obtained from just a handful
of response functions that represent certain contributions to the
triple commutator in Eq. (\ref{eq:rho3}). The details of the derivation
can be obtained e.g. in Ref. \cite{BrixnerMancal}. 

If the delays between the pulses are selected such that $\tau$ denotes
the delay between the first ($\bm{k}_{1}$) and the second ($\bm{k}_{2}$)
pulses, and $T$ denotes the delay between the second and third ($\bm{k}_{3}$)
pulse (e.g. $t_{3}=0$, $t_{2}=-T$ and $t_{1}=-T-\tau$) we can write
for the time and the delay dependent signal field\[
E_{s}(t,T,\tau)\approx R_{2g}(t,T,\tau)\]
\begin{equation}
+R_{3g}(t,T,\tau)-R_{1f}^{*}(t,T,\tau),\ \tau\geqq0,\label{eq:Es_reph}\end{equation}
\[
E_{s}(t,T,\tau)\approx R_{1g}(t,T,|\tau|)\]
\begin{equation}
+R_{4g}(t,T,|\tau|)-R_{2f}^{*}(t,T,|\tau|),\ \tau<0.\label{eq:Es_nonreph}\end{equation}
The absolute value in Eq. (\ref{eq:Es_nonreph}) originates from the
fact that response functions $R$ are defined for positive time arguments
only, and negative $\tau$ is achieved by switching the order of the
$\bm{k}_{1}$ and $\bm{k}_{2}$ pulses. The individual response functions
$R$ are listed in the Appendix \ref{sec:App_Third_order}. Most importantly,
they consist of series of propagation of the density matrix blocks
by evolution operators obtained from the solution of equations of
motion. We have e.g.\[
R_{2g}(t,T,\tau)=tr\{\mu_{ge}{\cal U}_{egeg}(t){\cal V}_{eg}^{(R)}\]
\begin{equation}
\times{\cal U}_{eeee}(T)V_{eg}^{(L)}{\cal U}_{gege}(\tau)V_{ge}^{(R)}\rho_{0}\},\label{eq:R2g_text}\end{equation}
where the evolution superoperators ${\cal V}_{ab}^{(R)}$ act on an
arbitrary operator $A$ as a dipole operator $\mu_{ab}$ from the
right, i.e. ${\cal V}_{ab}^{(R)}A=A\mu_{ab}$. The superoperator ${\cal V}_{ab}^{(L)}$
is defined analogically with the action of $\mu_{ab}$ from the left.
The indices $e$ and $g$ denote electronic bands as denoted in Fig.
\ref{fig:Exciton_states}. Thus, the above operators and the action
of superoperators on an arbitrary operator $A$ are expressed in the
basis of Hamiltonian eigenstates as\begin{equation}
\rho_{0}=|g\rangle\langle g|,\label{eq:rho0}\end{equation}
\begin{equation}
\mu_{eg}=\sum_{n}\mu_{ng}^{(eg)}|u_{n}\rangle\langle g|,\label{eq:mueg}\end{equation}
\begin{equation}
{\cal U}_{gege}(t)[A]=\sum_{nm}U_{gngm}^{(gege)}(t)\langle u_{m}|A|g\rangle|g\rangle\langle u_{n}|,\label{eq:Ugege}\end{equation}
\[
{\cal U}_{eeee}(t)[A]=\sum_{nn^{\prime}mm^{\prime}}U_{nn^{\prime}mm^{\prime}}^{(gege)}(t)\]
\begin{equation}
\times\langle u_{m^{\prime}}|A|u_{m}\rangle|u_{n}\rangle\langle u_{n^{\prime}}|.\label{eq:Ueeee}\end{equation}
Eqs. (\ref{eq:rho0}) to (\ref{eq:Ueeee}) together with the Appendix
\ref{sec:App_Third_order} enable us to calculate expected non-linear
signal from the knowledge of the matrix elements of the evolution
superoperator. This type of knowledge can be obtained from solutions
of the four different equations of motion that we presented in Section
\ref{sec:Second-order-relaxation}.

\subsection{Two-dimensional Coherent Spectroscopy\label{sub:Two-dimensional-Coherent-Spectroscopy}}

Two-dimensional coherent spectrum, $\Xi(\omega_{t},T,\omega_{\tau})$,
is obtain from the non-linear signal by Fourier transforming the time
and pulse delay dependent signal electric field $E_{S}(t,T,\tau)$
along the $t$ and $\tau$ variables \cite{Jonas01,BrixnerMancal}
as \begin{equation}
\Xi(\omega_{t},T,\omega_{\tau})=\int\limits _{-\infty}^{\infty}dt\int\limits _{-\infty}^{\infty}d\tau E_{s}(t,T,\tau)e^{i\omega_{t}t-i\omega_{\tau}\tau}.\label{eq:2D_spec}\end{equation}
The Fourier transform in $\tau$ yields an $\omega_{\tau}$ dependence
that is formally similar to linear absorption spectrum, while the
transform in $t$ yields generalized absorption and stimulated emission
from a non-equilibrium state created by the first two laser pulses.
2D spectrum thus represents a 2D absorption/emission and absorption/absorption
correlation plot. During the pulse delay time $T$ the system evolves
both in the electronically excited state and in the ground state,
but no optical signal is generated. Relaxation of populations in the
electronically excited band leads to evolution of non-diagonal 2D
spectral features, so-called cross-peaks. Cross-peaks appearing at
$T=0$ are a signature of excitonic origin of the observed excited
states. The 2D cross-peaks oscillate in $T$ as long as the corresponding
electronic coherence elements of the reduced density matrix are oscillating.
The life time of the electronic coherences can thus be estimated directly
from the $T$ dependent sequence of 2D spectra \cite{PisliakovMancal,EngelNature}.

\section{Numerical Results and Discussion\label{sec:Numerical-Results}}

In this section we study dynamics of model aggregate viewed via population
and coherence dynamics and via 2D coherent spectrum. We define a simple
model aggregate for which we calculate excited state dynamics including
evolution of coherences between electronic states, linear absorption
and 2D spectra at chosen population times. Calculations of linear
absorption, which require only knowledge of the time evolution of
optical coherences, are performed using the secular time local equation,
since it is known to yield exact result at least for some models \cite{Haenggi08}.
Population dynamics is calculated using all four methods we discussed
in Section \ref{sec:Second-order-relaxation}, and the results are
compared.

\begin{table}
\begin{tabular}{|c|c|c|c|c|c|c|c|}
\hline 
$n$ & $\frac{\epsilon_{n}}{cm^{-1}}$ & $\frac{d_{n,x}}{|d_{n}|}$ & $\frac{d_{n,y}}{|d_{n}|}$ & $\frac{d_{n,z}}{|d_{n}|}$ & $\frac{|d_{n}|}{d_{0}}$ & $h_{n}$ & $\frac{\alpha_{n}}{grad}$\tabularnewline
\hline 
1 & $9850$ & $1$ & $0$ & $0$ & $0.65$ & $0$ & $0$\tabularnewline
\hline 
2 & $10000$ & $-0.94$ & $0.34$ & $0$ & $2.15$ & $10$ & $60$\tabularnewline
\hline 
3 & $10150$ & $-0.94$ & $0.34$ & $0$ & $0.9$ & $10$ & $120$\tabularnewline
\hline
\end{tabular}

\caption{\label{tab:Model_Ia}Parameters of the model trimer. The parameter
$\epsilon_{n}$ represents the transition energy of $n-$th monomer,
transition dipole moments $\bm{d}_{n}$ are taken relative to some
value $d_{0}$. Parameters $h_{n}$ and $\alpha_{n}$ are explained
on Fig. \ref{fig:Trimer_Geometry}. }

\end{table}

\begin{figure}
\includegraphics[width=0.9\columnwidth]{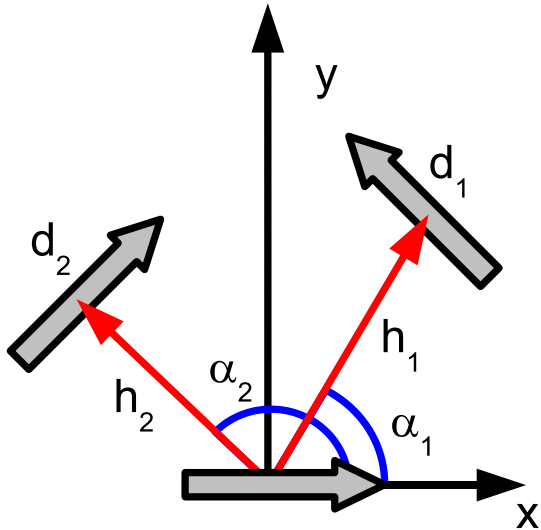}

\caption{\label{fig:Trimer_Geometry}Geometry and parameters of a trimer aggregate.
One monomer is chosen to be positioned at the origin of the coordinate
system, with the transition dipole moment pointing along the $x$
axis. The positions the transition dipole moments of the other two
molecules in space are characterized by their distance $h_{2}$ and
$h_{3}$ from the origin of coordinates and by the angles $\alpha_{2}$
and $\alpha_{3}$. Orientations and lengths of the dipoles are given
in Tab. \ref{tab:Model_Ia}. In our example we assume that the aggregate
is planar.}

\end{figure}

\begin{figure}
\includegraphics[width=1\columnwidth]{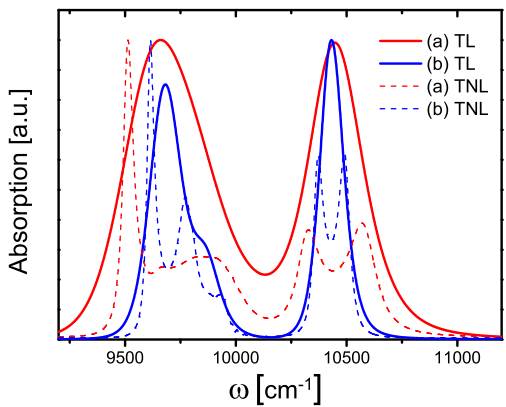}

\caption{\label{fig:Linear-absorption-spectrum}Linear absorption spectrum
of the model trimer for various parameters of the system bath interaction:
(a) $\lambda=120$ cm$^{-1}$, $\tau_{c}=50$ fs, (b) $\lambda=30$
cm$^{-1}$ , $\tau_{c}=100$ fs, calculated by the secular TL theory
(full lines) and the secular TNL theory (dashed lines). }

\end{figure}

The simplest model of an aggregate that can exhibit all effects observed
in Ref. \cite{EngelNature} is a trimer. The geometry of the models,
together with the meaning of the parameters is presented in Fig. \ref{fig:Trimer_Geometry}.
In Tab. \ref{tab:Model_Ia} we summarize the main parameters of the
model. Parameters $J$, $h$ and $d_{0}$ from Tab. \ref{tab:Model_Ia}
are not independent. For given $h_{n}$ and $\bm{d}_{n}$ we could
in principle calculate the value of resonance coupling $J$. Because
we are not interested in the absolute amplitude of the absorption
or 2D spectra we assume $d_{0}$ to be fixed by the values of $h_{n}$
and $\bm{d}_{n}$ to yield the expected value of $J$. All three resonance
couplings $J$ between the molecules are set to $J=200$ cm$^{-1}$
for the calculations presented here. The values of the transition
dipole moments determine the initial condition for the population
dynamics. We assume that the excitation light intensity and the value
of the transition dipole moment are such that the system is only weakly
excited. The total population of the excited state band is normalized
to $0.01$. The relative values of the transition dipole moments are
chosen so that the linear absorption spectrum (see Fig. \ref{fig:Linear-absorption-spectrum})
shows peaks of roughly the same height. Two peaks originating from
the energetically lowest and the energetically highest states dominate
the spectrum, the third level contributes as a shoulder to lowest
energy peak. 

Two parameters that influence the coupling for the model system to
the bath are reorganization energy $\lambda$ and correlation time
$\tau_{c}.$ We vary these parameters in the range that can conceivably
represent chlorophylls in photosynthetic complexes (see e. g. Refs.
\cite{Zigmantas,Cho}).

\subsection{Population relaxation and evolution of coherences\label{sub:Population-relaxation}}

\begin{figure}
\includegraphics[clip,width=1\columnwidth]{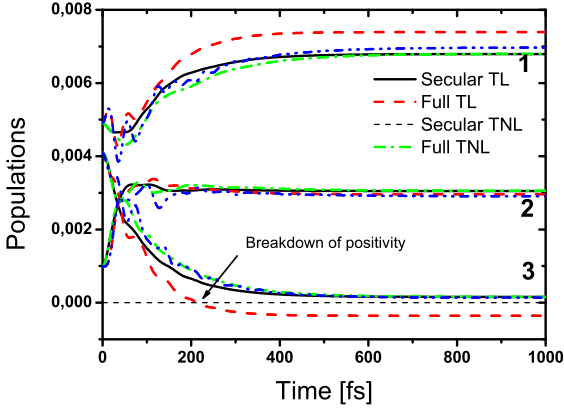}

\caption{\label{fig:population} First 1000 fs of the excited state population
dynamics of a trimer with parameters $\lambda=120$ cm$^{-1}$, $\tau_{c}=50$
fs, calculated by all four methods. For these particular parameters,
the full TL equation breaks positivity of the RDM diagonal elements
after $200$ fs. Its prediction for the populations of the lowest
and highest levels is significantly different from the other three
methods.}

\end{figure}

\begin{figure}
\includegraphics[clip,width=1\columnwidth]{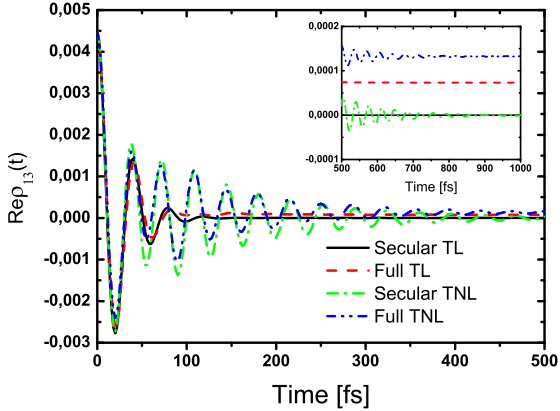}

\caption{\label{fig:coherences}First 500 fs of the dynamics of the RDM coherence
element $\rho_{13}(t)$, with parameters from Fig. \ref{fig:population},
calculated by all four methods. Detail of the long time part of the
time evolution is presented in the inset.}

\end{figure}

First, we compare relaxation dynamics of populations of excited state
of our aggregate after excitation by an ultrashort laser pulse. TL
equations of motion where solved by standard numerical methods for
ordinary differential equations provided by the Mathematica\textsuperscript{\textregistered}
software. For the TNL equations we used fast Fourier transform method.
Figure \ref{fig:population} presents the first $1$ ps of the population
dynamics after a $\delta-$pulse excitation of the trimer from Tab.
\ref{tab:Model_Ia} at the temperature ${\cal T}=300$ K. Reorganization
energy $\lambda=120$ cm$^{-1}$ and correlation time $\tau_{c}=50$
fs are the same at all three monomers. The dynamics with the same
parameters for a selected coherences element $\rho_{13}(t)$ is presented
in Fig. \ref{fig:coherences}. The overall conclusion is that all
four methods yield a similar general behavior for the populations,
with some difference at the short time evolution and also slightly
different long time equilibrium. Examination of the Figure \ref{fig:coherences}
leads us to the conclusion that the methods yield two different results
- a short coherence life time for the time local methods, and a relatively
longer life time in case of the time non-local methods. The behavior
of the coherence $\rho_{13}(t)$ represents a general tendency that
we have observed for all electronic coherences over a wide range of
parameters.

\begin{figure}
\includegraphics[width=1\columnwidth]{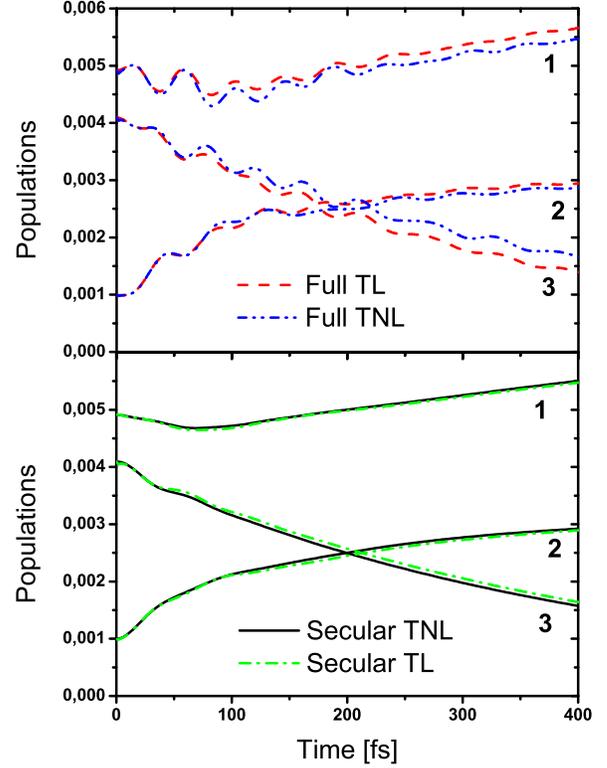}

\caption{\label{fig:Population-detail}First $400$ fs of the population dynamics
of the trimer with parameters $\lambda=30$ fs and $\tau_{c}=100$
fs. Results of full TL and TNL theories are presented in upper subfigure
(A), the secular results are found the the lower subfigure (B). }

\end{figure}

Let us now concentrate on short time behavior of the populations and
coherences in more retail. In the short time evolution of the coherences
the four methods group into two distinct groups with short (TL methods)
and long (TNL methods) coherence life time. Whether the underlying
equation is secular or not seems to have only a little influence on
the coherence dynamics. Fig. \ref{fig:Population-detail} shows the
short time ($0-400$ fs) comparison of the population calculated by
the four equations of motion. We can clearly see that the results
can be naturally grouped according to the presence of fast oscillatory
modulation of the population relaxation dynamics. In the one group
we have the full TL and full TNL methods, where such oscillations
clearly occur, the second group comprises the two secular methods
with no oscillations present. Thus, it can be concluded that the non-secular
terms in the equations of motion are the cause of these oscillations.
This is also supported by comparison of the population dynamics of
the full TL and full TNL equations from Fig. \ref{fig:population}
(e.g. the population of the state $1$). The oscillation on the full
TNL curve last longer than those of the full TL one, which reflects
the longer coherence life time we have found for the TNL equations. 

Let us now discuss the long time limit of the time evolution. As expected,
the two secular theories yield the same equilibrium at long population
times. This equilibrium corresponds to the canonical distribution
of population among the excitonic levels at ${\cal T}=300$ K. In
both secular TNL and secular TL cases, coherences have relaxed to
zero at long times as the inset of the Fig. \ref{fig:coherences}
demonstrates. The non-secular TNL and TL equations yield non-zero,
stationary coherences at long times, and correspondingly, the long
time equilibrium populations do not correspond to the canonical thermal
equilibrium. Although both non-secular theories converge to results
different from the canonical equilibrium, the full TNL equation yields
populations that are physical at all times for the studied system
parameters, i.e. they are always positive. The full TL equation on
the other hand fails to keep probabilities positive at long times,
and the occupation probability of the highest electronic level becomes
negative after $200$ fs for the parameters used on Fig. \ref{fig:population}.

\begin{figure}
\includegraphics[clip,width=1\columnwidth]{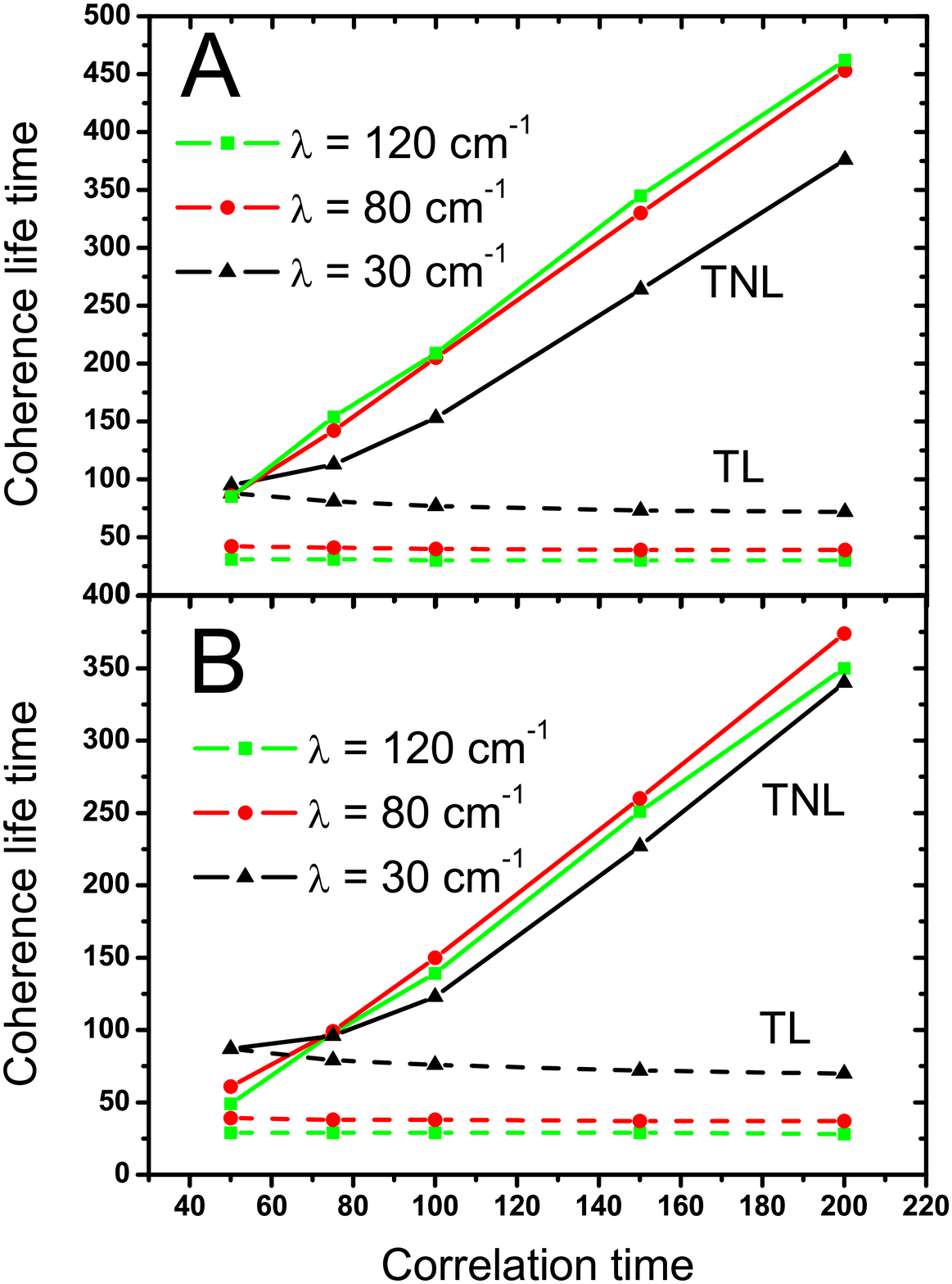}

\caption{\label{fig:lifetimes}The life time of coherence $\rho_{13}(t)$ as
obtained from fitting the coherence dynamics calculated by all four
methods for various parameters $\lambda$ and $\tau_{c}$. The upper
subfigure (A) shows the life times obtained by the secular methods,
while the lower subfigure (B) presents the same for non-secular methods.}

\end{figure}

In light of recent experiments \cite{EngelNature}, the conclusion
that time non-local theories lead to a longer coherence life time
than the time-local ones (i.e. also longer than the standard constant
rate theories) is probably the most interesting. We have performed
calculations of the RDM dynamics while varying the reorganization
energy and the correlation time. The absolute values of the coherence
$\rho_{13}(t)$ elements were fitted by a single exponential to estimate
coherence life-time. The results are summarized in Fig. \ref{fig:lifetimes}.
The Fig. \ref{fig:lifetimes}A shows the results for secular TL and
secular TNL equations. Clearly, with growing correlation time $\tau_{c}$,
the full TNL equations lead to a increasing coherence life time. The
full TL equation shows only a very weak dependence of the coherence
life time on correlation time. Another interesting observation is
that for correlation time longer then $50$ fs, the dependence of
the coherence life time on the reorganization energy $\lambda$ is
different for full TNL and TL methods. Time local theory, in accordance
with the standard rate theories, predicts decrease of the coherence
life time with $\lambda$. The full TNL theory predicts (within the
parameter range studied here) an opposite tendency. The Fig. \ref{fig:lifetimes}B
shows similar conclusion for the non-secular versions of the theories,
with the same difference between TL and TNL theory. The dependence
of the coherence life time on $\lambda$ in case of TNL equations
is not monotonous. 

\subsection{Two-dimensional spectrum}

As discussed in the Introduction, the secular TL equation of motion
yields an exact result for the dephasing of an isolated optical coherence
\cite{Haenggi08}. One can show, by comparison of the absorption spectra
calculated by secular TL and TNL methods (see Fig. \ref{fig:Linear-absorption-spectrum}),
that the TNL theory leads to certain artifacts (second peak) and is
therefore not suitable for the description of the optical coherence
evolution. Consequently, on can only hope to obtain valid results
for the evolution superoperators at the first and the third time interval
of the third order response functions by the TL theories. In Ref.
\cite{Mancal04a} it was shown that non-secular terms in the TL equations
for optical coherences lead to temperature dependence of the positions
of excitonic bands in absorption spectra. This dependence was shown
to be strong when the electronic states involved are characterized
by significantly different reorganization energy \cite{Mancal04a,Mancal08a}.
Indeed it can be shown for homodimer that the non-secular terms are
exactly zero in second order TL theory if the monomers exhibit the
same reorganization energy \cite{Mancal08a}. We can therefore expect
the non-secular effects in the optical coherences to be weak in our
case, and we choose secular TL to calculate the evolution superoperators
in the first and third time interval of the response function, Eq.
(\ref{eq:R2g_text}). 

\begin{figure}[H]
\includegraphics[clip,width=1\columnwidth]{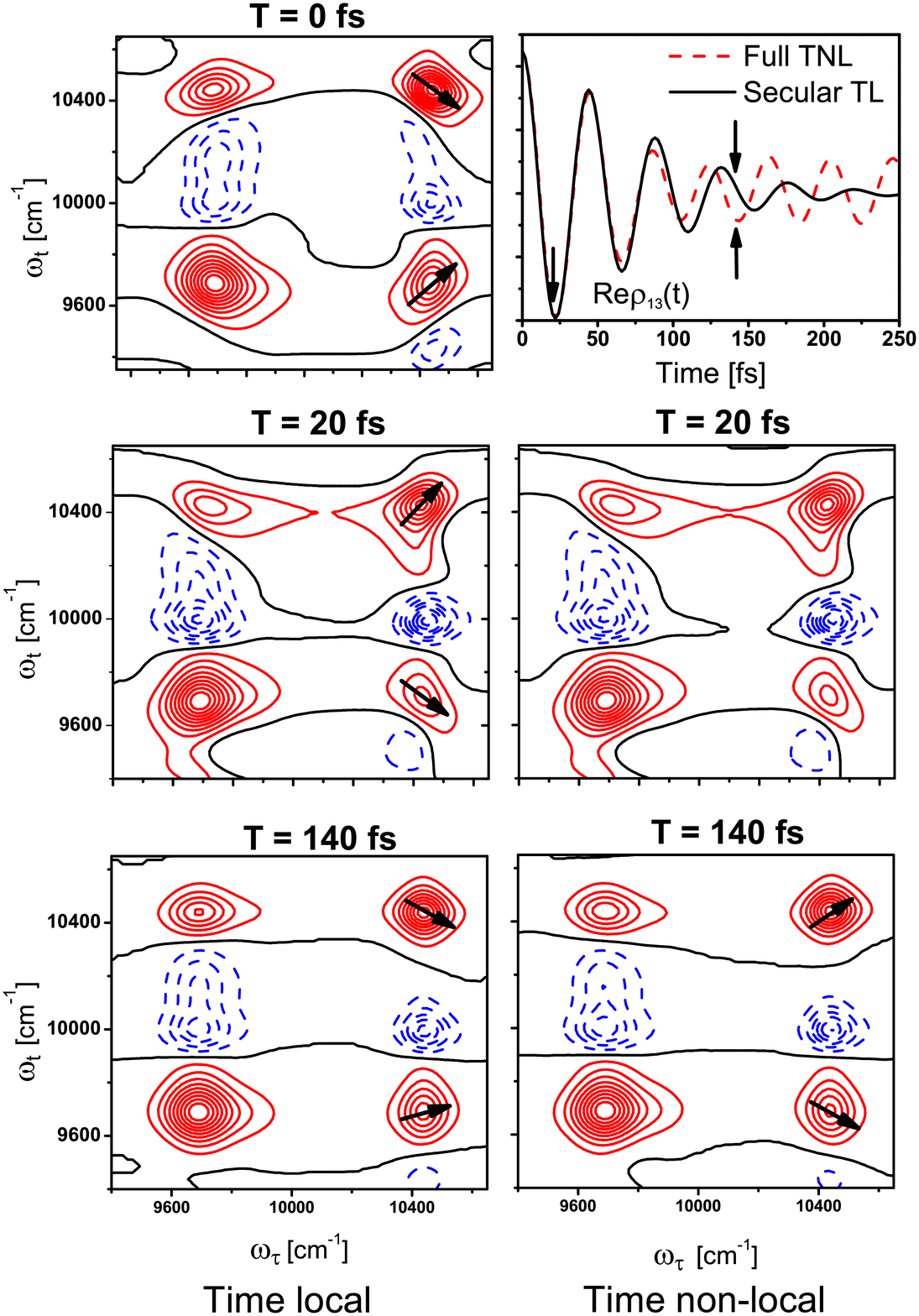}

\caption{\label{fig:2Dfig}Two-dimensional coherent spectra of the trimer model
at population times $T=0$, $20$ and $140$ fs calculated by the
secular TL method (left column) and the full TNL method (right column).
The system-bath interaction parameters are $\lambda=30$ cm$^{-1}$
and $\tau_{c}=100$ fs. The coherence element $\rho_{13}(t)$, which
is mainly responsible for the oscillatory behavior of the crosspeaks,
is presented in the upper right corner of the figure. The 2D spectrum
at $T=0$ fs is the same for both methods and is therefore presented
only once. The population times are selected so that they represent
different phases of the $\rho_{13}(t)$ element (denoted by arrows
on the coherence element figure). Arrows in the 2D spectra denote
the orientation of the peaks. All spectra are normalized to 1 with
contour step of $10$ \%. Positive features are in full red line,
negative features are represented by dashed blue line, and the zero
contour is depicted by the full black line.}

\end{figure}

Concerning the population interval, the situation is somewhat different.
As we have shown above, the non-secular TL theory leads to dynamics
that breaks the positivity condition for the population probabilities
at long times. At the same time, short time dynamics is very similar
to the full TNL. Both theories predict population oscillations during
the life time of the electronic coherences. The full TNL equation,
however, preserves positivity, at least for the parameters studied
here, and can be therefore used to calculate meaningful 2D spectra.
For the same reason, both secular theories can also be successfully
used to calculate 2D spectrum. As the oscillation of the populations
predicted by non-secular theories are too small to be reliably observed
in 2D spectrum (only a small change of the crosspeak amplitude due
to the population transfer is observed after $140$ fs of relaxation
in 2D spectrum of Fig. \ref{fig:2Dfig} ) we expect only a small difference
of the 2D spectrum to appear between the secular and full TNL theories.
For the calculation of the representative 2D spectrum we therefore
choose the secular TL and the full TNL theories. These two differ
from each other mainly in the prediction of the life time of the electronic
coherences. The observable difference in the calculated 2D spectra
should therefore predominantly result from the different life time
of the electronic coherence. 

Fig. \ref{fig:2Dfig} presents 2D spectra for $\lambda=30$ cm$^{-1}$
and $\tau_{c}=100$ fs. These parameters lead to a rather slow relaxation
and consequently to narrow spectral peaks in both absorption (see
Fig. \ref{fig:Linear-absorption-spectrum}) and 2D spectra. This allows
us to clearly see characteristic $T-$dependent oscillations of the
peaks in 2D spectrum. At $T=0$ fs, both methods provide the same
2D spectrum, with four peaks. Two diagonal peaks arise when all three
perturbations of the system by electric field occur on the same level,
while two crosspeaks appear from interactions occurring on different
levels. Negative peaks correspond to excited state absorption (see
Fig. \ref{fig:Exciton_states}). For two molecules that are not excitonically
coupled, all contributions to the crosspeaks cancel out exactly, while
if two molecules are excitonically coupled non-zero crosspeaks appear.
The shapes of the peaks are influenced by the phase evolution of the
coherence elements of RDM during the population time $T$. On the
upper left figure of Fig. \ref{fig:2Dfig} we have marked the elongation
of the diagonal and off-diagonal peaks by arrows. The elongation can
be best judged by looking at the zero contour (in black). This particular
elongation is characteristic for the phase of the $\rho_{13}(t)$
element (see upper right figure of Fig. \ref{fig:2Dfig}) at $T=0$.
At $T=20$ fs the phases of the $\rho_{13}(t)$ calculated by both
methods are opposite to the phase at $T=0$. The 2D spectra calculated
by the two different methods at $T=20$ fs differ only in the precise
positions of the contours. This phase of the coherence element is
characterized in 2D spectrum by a different orientation of the peaks.
Interestingly, at $T=140$ fs the two methods predict $\rho_{13}(t)$
that have mutually opposite phases and as a consequence the 2D spectra
at $T=140$ fs calculated by different methods differ in the orientation
of their crosspeaks. Since the secular TL theory predicts a simple
dephasing of the coherence and a regular oscillation with a single
frequency proportional to the energy difference between corresponding
energy levels, it is in principle possible to distinguish, even experimentally,
deviations from this prediction. Our conclusion is that such a deviation
should be a consequence of the memory effects in the reduced system
time evolution.

\subsection{Validity of secular and Markov approximations}

Several conclusions about the applicability of the secular and Markov
approximations can be drawn from the above results. As pointed out
in Ref. \cite{Haenggi08}, Markov approximation, which in the second
order in system-bath coupling converts the TNL equations to the TL
ones, leads accidentally to an exact result for an optical coherence
element interacting with the harmonic bath. It has been also pointed
out previously \cite{Ishizaki2,Kubo69} that in the same case, the
TNL equations lead to artifacts. When studying relaxation dynamics
of the populations and electronic coherences in excitonic systems,
full TL theory leads to a breakdown of the positivity of the RDM,
while none of the secular theories suffer from this problem. In principle,
the full TNL theory suffers from this problem, too. However, it has
been found less susceptible to it here. The secular theories lead
to canonical density matrix at long times, while the full TNL results
in a stationary state characterized by non-zero (but constant) coherences.
Such result corresponds to an additional renormalization of the electronic
states by the interaction with bath, and has to be expected even at
a weak coupling limit \cite{Tannor00}. It is important to note in
this context that the canonical equilibrium is to be expected for
the system consisting of the molecule and the bath as a whole, not
for its parts \cite{Tannor00}. 

For the population dynamics we are therefore forced to conclude that
the full TNL theory represents the best candidate for a correct description
of relaxation phenomena in the second order of the system bath interaction.
It predicts similar population transfer times as other methods, it
is much less sensitive to the breakdown of the positivity than its
TL counterpart, and it leads to a bath renormalization of the canonical
equilibrium. Most interestingly however, it predicts longer coherence
life time than the TL theory. It was recently established by Ishizaki
and Fleming \cite{Ishizaki2} that this is to be expected from a higher
order theory. 

In the light of the above conclusions about the dynamics of optical
coherences and the populations and coherences of the one exciton band,
we suggest a hybrid approach to calculating 2D spectra, which consists
of the application of the TL method on optical coherences (first and
third time interval) and the full TNL method on the calculation of
the RDM dynamics in the one exciton band during the population time
$T$.

\section{Conclusions\label{sec:Conclusions}}

In this paper we have compared four different theories of excitation
energy transfer and relaxation in molecular aggregate systems, with
a special attention paid to lifetime of electronic coherences. Second
order time non-local and time local theories with and without secular
approximation were studied. For our specific model of an aggregate
we have concluded that time non-local theories can account for experimentally
observed electronic coherence life time that is significantly longer
than the one predicted by the standard time-local secular relaxation
rate theory. Markov approximation leading to time local equations
of motion was found to be responsible for the reduction of the coherence
lifetime, while the influence of the secular approximation on the
life time was found rather weak. The time local theory without secular
approximation is found to break positivity of the occupation probabilities
in the range of parameters studied here. We conclude that time-local
second order theory is not suitable for simulating the coherence transfer
effects. Simulations of two-dimensional spectra show that the time
non-local effects can be experimentally identified based on the analysis
of the oscillations of the cross peaks. 
\begin{acknowledgments}
This work was supported by the grant KONTAKT ME899 from the Ministry
of Education, Youth and Sports of the Czech Republic. Two-dimensional
spectra were produced using the NOSE package available under GNU Public
License at \emph{http://www.sourceforge.net}.
\end{acknowledgments}
\appendix

\section{Third Order Response Functions\label{sec:App_Third_order}}

In this appendix we list the third order response function used in
calculating the impulsive 2D spectra. The first index of the response
function follows the standard notation of Ref. \cite{MukamelBook}.
The second index is $g$ for pathways not involving the two-exciton
band, while all pathways denoted by $f$ include a two-exciton contribution
(see e.g. Ref. \cite{BrixnerMancal}). \[
R_{1g}(t,T,\tau)=tr\{\mu_{ge}{\cal U}_{egeg}(t){\cal V}_{eg}^{(R)}\]

\begin{equation}
\times{\cal U}_{eeee}(T){\cal V}_{ge}^{(R)}{\cal U}_{egeg}(\tau){\cal V}_{eg}^{(L)}\rho_{0}\},\label{eq:R1g_app}\end{equation}
\[
R_{2g}(t,T,\tau)=tr\{\mu_{ge}{\cal U}_{egeg}(t){\cal {\cal V}}_{eg}^{(R)}\]
\begin{equation}
\times{\cal U}_{eeee}(T){\cal V}_{eg}^{(L)}{\cal U}_{gege}(\tau){\cal V}_{ge}^{(R)}\rho_{0}\},\label{eq:R2g_app}\end{equation}
\[
R_{3g}(t,T,\tau)=tr\{\mu_{ge}{\cal U}_{egeg}(t){\cal V}_{eg}^{(L)}\]
\begin{equation}
\times{\cal U}_{gggg}(T){\cal V}_{eg}^{(R)}{\cal U}_{gege}(\tau){\cal V}_{ge}^{(R)}\rho_{0}\},\label{eq:R3g_app}\end{equation}
\[
R_{4g}(t,T,\tau)=tr\{\mu_{ge}{\cal U}_{egeg}(t){\cal V}_{eg}^{(L)}\]
\begin{equation}
\times{\cal U}_{gggg}(T){\cal V}_{ge}^{(L)}{\cal U}_{egeg}(\tau){\cal V}_{eg}^{(L)}\rho_{0}\},\label{eq:R4b_app}\end{equation}
\[
R_{1f}(t,T,\tau)=tr\{\mu_{fe}{\cal U}_{efef}(t){\cal V}_{ef}^{(R)}\]
\begin{equation}
\times{\cal U}_{eeee}(T){\cal V}_{ge}^{(R)}{\cal U}_{egeg}(\tau){\cal V}_{eg}^{(L)}\rho_{0}\}\label{eq:R1f_app}\end{equation}
\[
R_{2f}(t,T,\tau)=tr\{\mu_{fe}{\cal U}_{efef}(t){\cal V}_{ef}^{(R)}\]
\begin{equation}
\times{\cal U}_{eeee}(T){\cal V}_{eg}^{(L)}{\cal U}_{gege}(\tau){\cal V}_{ge}^{(R)}\rho_{0}\}.\label{eq:R2f_app}\end{equation}
Operators and superoperators used in this appendix are defined in
Section \ref{sub:Third-order-non-linear-response}.

\bibliographystyle{prsty}
\bibliography{OlsinaMancal_arXiv_2009}

\end{document}